\documentclass[11pt]{article}
\usepackage{amssymb}
\usepackage{multicol}
\usepackage{amsmath}
\usepackage{longtable}

\newtheorem{example}{Example}[section]

\newtheorem{note}{Note}[section]

\begin{document}
\today
\begin{center} 
\begin{Large}
{\LARGE\bf  
The distribution of the maximum of a first order
moving average: the discrete case
}\\[1ex]
\end{Large}
by\\[1ex]
Christopher S. Withers
\\
Applied Mathematics Group\\
Industrial Research Limited\\
Lower Hutt, NEW ZEALAND
\\[2ex] Saralees Nadarajah
\\ School of Mathematics\\ University of Manchester\\ Manchester M60 1QD, UK
\end{center}
\vspace{1.5cm}
{\bf Abstract:}~~We give the distribution of $M_n$,  the maximum of a
sequence of $n$ observations from a moving average of order 1. 
Solutions are first given in terms of repeated integrals
and then for the case where the underlying independent random variables
are discrete.
A solution appropriate for large $n$ takes the form
$$
Prob(M_n 
 \leq x)\ =\ \sum_{j=1}^I \beta_{jx} \ \nu_{jx}^{n}
\ \approx \ B_{x}\ r_{1x}^{n}
$$
where 
$\{\nu_{jx}\}$ are the eigenvalues  of a certain matrix, $r_{1x}$ is the 
 maximum magnitude of the eigenvalues, and $I$ depends on the number of 
possible values of the underlying random variables.
The eigenvalues do not depend on $x$ only on its
range. 

\section{Introduction and summary}
\setcounter{equation}{0}
We give the distribution of the maximum of a moving average of order 1 for
discrete random variables.

Section 2 summarises results for any moving average of order 1 (discrete or
not) given in Withers and Nadarajah (2009). Two forms are given for 
 the distribution of the maximum. Only one of these is appropriate for large
$n$. This form can be viewed as a large deviation expansion.
It assumes that a related parameter $v_n$ can be written as a weighted sum
of $n$th powers.

Section 3 gives 3 sets of situation of increasing generality where this last
assumption holds. 

Let  $\{e_i\}$ be independent and identically distributed random variables
 from some distribution $F$ on $R$.
Consider the moving average of order 1,
\begin{eqnarray*}
X_i= e_i+ \rho e_{i-1} 
\end{eqnarray*}
where $\rho\neq 0.$
So the observations take the values 
$\{x_i+\rho x_j\}$.
In Withers and Nadarajah (2009) we gave
 expressions for the distribution of the maximum
$$M_n=\max_{i=1}^n X_i$$
in terms of repeated integrals.
This was obtained via the recurrence relationship
\begin{eqnarray}
G_{n}(y) &=& I(\rho<0)G_{n-1}(\infty)F(y) +  {\cal K}  G_{n-1}(y)
 \label{joint} \\
\mbox{where }G_{n}(y) &=& Prob(M_n\leq x,\ e_n\leq y), \label{defG}
\end{eqnarray}
 $I(A)=1$ or 0 for $A$ true or false,
and
 $ {\cal K}$ is the  integral operator
\begin{eqnarray}
{\cal K} r(y) = \mbox{sign}(\rho) \int^y r((x-w)/\rho) dF(w).\label{calK}
\end{eqnarray}
For this to work at $n=1$, define
$$M_0=-\infty\mbox{ so  that  }G_{0}(y)=F(y).$$
Note that {\it dependence on $x$ is suppressed.}

Our purpose is to find
\begin{eqnarray*}
 u_n=Prob(M_n\leq x)= G_{n}(\infty),\ n\geq 0. 
\end{eqnarray*}
Section 2 summarises and extends relevant results in   Withers and Nadarajah 
(2009).

In Section 3 we consider the case where $e_1$ is discrete and derive a solution
of the form given in the abstract.

Section 4 gives a solution to $G_n$ of (\ref{defG}).
For any integrable function $r$, set 
$\int r=\int r(y)dy=\int_{-\infty}^\infty r(y)dy,\
\int^x r=\int^x r(y)dy=\int_{-\infty}^x r(y)dy
$.

\section{Solutions using repeated integrals and sums of powers.}
\setcounter{equation}{0}
Set
\begin{eqnarray}
 v_n= [{\cal K}^n  F(y)]_{y=\infty}. \label{vn0} 
\end{eqnarray}
For example 
\begin{eqnarray}
v_0=1,\ v_1=-\int F(z)dF(x-\rho z) = -I(\rho<0)+\int F(x-\rho z) d F(z).\label{v1}
\end{eqnarray}
{\it The case $\rho>0$.} 
(\ref{joint}) has solution
\begin{eqnarray}
G_n(y)={\cal K}^n  F(y),\ n\geq 0, \label{Gpos}
\end{eqnarray}
so that
\begin{eqnarray}
 u_n=v_n,\ n\geq 0. \label{pos}
\end{eqnarray}
For example $u_0=1$.
(The marginal distribution of $X_1$ is $u_1=v_1$ given by (\ref{v1}).)

{\it The case $\rho<0$.} 
By (\ref{joint}) for $n\geq 0,$
\begin{eqnarray}
G_{n+1}(y) &=& u_{n}F(y)+  {\cal K}G_{n}(y)\nonumber \\
 &=& a_n(y) \otimes u_n +a_{n+1}(y)\label{2.4a}
\\ \mbox{where }
a_i(y) &=& {\cal K}^{i}F(y), \ a_n\otimes b_n=\sum_{j=0}^n a_jb_{n-j}. 
\label{2.4b} 
\end{eqnarray}
Putting $y=\infty$, $u_n$ is given by the recurrence equation
\begin{eqnarray}
 u_0=v_0=1,\ u_{n+1}=v_{n+1} +u_n\otimes v_n
,\ n\geq 0. \label{rec} 
\end{eqnarray}
(The marginal distribution of $X_1$ is $u_1=1+v_1$ of (\ref{v1}).)
The solution of (\ref{rec}) is
\begin{eqnarray}
u_{n}=\hat{B}_{n+1}(w),\ n\geq 0 \mbox{ where }w_n=v_{n-1}\label{w2u} 
\end{eqnarray}
and $\hat{B}_{n}(w)$ is the {\it complete ordinary Bell polynomial},
a function of $(w_1,\cdots,w_n)$ generated by
$$(1-\sum_{n=1}^\infty w_nt^n)^{-1}=\sum_{n=0}^\infty \hat{B}_{n}(w)t^n.$$
So $\{u_{n-1}\}$ have generating function
\begin{eqnarray}
tU(t)=(1-tV(t))^{-1}-1\mbox{ where }
U(t)=\sum_{n=0}^\infty u_nt^n, \ V(t)=\sum_{n=0}^\infty v_nt^n.
 \label{*} 
\end{eqnarray}
 For example 
since $v_0=1$, reading from a table gives
\begin{eqnarray}
u_0 &=&1,\ u_1=v_1+1 \nonumber \\
u_2 &=&v_2+2v_1+1, \nonumber \\
u_3 &=& v_3+(2v_2+v_1^2)+3v_1+1, \nonumber\\
u_4 &=&v_4+(2v_3+2v_1v_2)+(3v_2+3v_1^2)+4v_1+1.\label{4}
\end{eqnarray}
For $u_n$ up to $n=9$ and more details on computing $\hat{B}_{n}(w)$,
 see  Withers and Nadarajah (2009).
Note that $1\geq u_n=Prob(M_n\leq x)\geq u_{n+1}\geq 0$ so that
\begin{eqnarray}
u_n=0\Rightarrow u_{n+1}=0.\label{**}
\end{eqnarray}
The solution (\ref{w2u})
gives no indication of the behaviour of $u_n$ for large $n$.
 In Section 3 we shall see that we can usually write $v_n$ as a weighted sum
of powers, say
\begin{eqnarray}
v_{n}= \sum_{j=1}^I \beta_j \nu_j^{n-1} 
\mbox{ for } n\geq n_0 \label{vn} 
\end{eqnarray}
where $1\leq I\leq \infty,\ n_0\geq 0$.
We call this the {\it weighted-sum-of-powers assumption.}
In this case 
 $u_n$ generally has the form
\begin{eqnarray}
u_n = \sum_{j=1}^{J} \gamma_j \delta_j^n \mbox{ for } n\geq \max(0,2n_0-1) 
 \mbox{ where }J\leq I'=I+n_0,
\label{un2}
\end{eqnarray}
and $\{\delta_j\}$ are 
the roots of
\begin{eqnarray}
&& \sum_{k=1}^I \beta_k\nu_k^{n_0-1}/(\delta-\nu_k)=p_{n_0}(\delta)
\mbox{ where } p_{n+1}(\delta)=\delta^{n+1}-v_{n}\otimes \delta^{n}:
\label{d} \\
&& p_0(\delta)=1 ,\ p_1(\delta)=\delta-1,\ 
p_2(\delta)=\delta^2-\delta-v_1,\
p_3(\delta)=\delta^3-\delta^2-\delta v_1-v_2,\ \cdots \nonumber
\end{eqnarray}
So (\ref{d}) can be written as a polynomial in $\delta$ of degree $J$
where $J\leq I'$.
\\ {\it Case 1:} Assume that these  $J$ roots are all distinct.

Having found  $\{\delta_j\}$ ,  $\{\gamma_j\}$ are the roots of
\begin{eqnarray}
&&\sum_{j=1}^{J}A_{jn_0}(\nu)\gamma_j =q_{n_0}(\nu)
\mbox{ for }\nu=\nu_1,\cdots,\nu_I\label{g} \\
\mbox{where} && A_{jn}(\nu)=\delta_j^{n}/(\delta_j-\nu),\
q_{n+1}(\nu)=\nu^{n+1} +u_{n}\otimes \delta^{n}:\nonumber\\
&& q_{0}(\nu)=1,\ q_{1}(\nu)=\nu+1,\  q_{2}(\nu)=\nu^2+\nu+u_1,\  
q_{3}(\nu)=\nu^3+\nu^2+u_1\nu+u_2,\ \cdots\nonumber
\end{eqnarray}
(\ref{g}) can be written 
\begin{eqnarray}
A_{n_0}{\bf \gamma}=Q_{n_0}\mbox{ where } (A_n)_{kj}=A_{jn}(\nu_k),
\ Q_{n}=(Q_{n1},\cdots, Q_{nI'})',
\ Q_{nk}=q_n(\nu_k).\label{Q0}
\end{eqnarray}
So  if $J=I$, a solution is
\begin{eqnarray}
{\bf \gamma}=A_{n_0}^{-1} Q_{n_0}. \label{gam}
\end{eqnarray}
(If $I=\infty$, numerical solutions can be found by truncating the infinite matrix $A_n$ 
and infinite vectors $( Q_n, \gamma)$
to $N\times N$ matrix and $N$-vectors, then increasing $N$ until the 
desired precision is reached.)
The proof, which is by substitution, assumes that $\{\delta_j,\nu_j\}$ are all distinct. The proof relies on the fact that if $\sum_{j=1}^J a_jr_j^n=0$
for $1\leq n \leq J$ and $r_1,\cdots, r_J$ are distinct, then $a_1=\cdots=a_J=0$, since $\det(r_j^n:\ 1\leq n,j \leq J)\neq 0.$ (\ref{un2}) 
 extends a corresponding result in  Withers and Nadarajah (2009a).
(If $J<I$, a solution is given by dropping $I-J$ rows of (\ref{Q0}).
If $J>I$, there are not enough equations for a solution, and the 2nd method 
below needs to be used.)
 
The values of $u_n$ for $n<2n_0-1$ can be found from (\ref{rec}) or (\ref{w2u})
or the extension of (\ref{4}).

{\it Behaviour for large $n$.}
If (\ref{vn}) holds then
$$v_n\approx B r_1^n \mbox{ as }n\rightarrow \infty
\mbox{ where }B=\sum_j \{ \beta_j\nu_j^{-1} e^{i\theta_j n}: |\nu_j|=r_1\},
$$
$r_1=\max_{j=1}^I |\nu_j|,\ \nu_j=r_je^{i\theta_j }$.
A similar result holds for $u_n$ of (\ref{un2}).
 Withers and Nadarajah (2009b) give a 2nd method of solution based on
 (\ref{*}), that applies even when the weights $\beta_j$ in (\ref{vn}) are 
polynomials in $n$.

\section{The discrete case.}
\setcounter{equation}{0}

Suppose that $e_1$ is a discrete random variable, say
$$e_1=x_i \mbox{ with probability } p_i>0 \mbox{ for } i=1,2,\cdots,P
\mbox{ where }1\leq P\leq \infty$$
and $ \sum_{i=1}^P p_i=1.$ 
We do not need to assume that $x_1<x_2<\cdots.$
(The method extends in an obvious way for $i=\cdots,-1,0,1,2,\cdots $
where $ \sum_{i=-\infty}^\infty p_i=1.$)

So the observations $X_1,\cdots, X_n$ 
take their values from the lattice $\{x_i+\rho x_j,\ 1\leq i,j\leq P\}.$ 
Set $I(A)=1$ or 0 for $A$ true or false. 
The main task of this section is to give three increasingly general situations
where the weighted-sum-of-powers assumption (\ref{vn}) holds.
For any function $H$, set
\begin{eqnarray}
 A_{iH} &=& \mbox{sign}(\rho)\ p_i\ H((x-x_i)/\rho),
\ A_{H}= (A_{1H},\cdots,A_{PH})',
\nonumber \\
q_i(y) &=& I(x_i\leq y), \  q(y)=(q_1(y),\cdots, q_P(y))',\nonumber \\
Q_{ij} &=&\mbox{sign}(\rho)\ q_i((x-x_j)/\rho)\ p_j,\ 
Q=(Q_{ij}:1\leq \ i,j\leq P). \label{Q}
\end{eqnarray}
(Recall that dependence on $x$ is suppressed.)
For example
\begin{eqnarray*}
Q_{12}\ \mbox{sign}(\rho)
& =& I(x_1\leq (x-x_2)/\rho)\\
&=& 1\iff
 \rho>0, x\geq x_2+\rho x_1\mbox{ or } \rho<0, x\leq x_2+\rho x_1.
\end{eqnarray*}
  Note that 
\begin{eqnarray*}
{\cal K}H(y) =  A_{H}' q(y),\
{\cal K}q(y) = Q  q(y),\
{\cal K}^{n}H(y) =  A_{H}'Q^{n-1} q(y) \mbox{ for } n\geq 1.
\end{eqnarray*}
So $v_0=1$ and for $n\geq 1$, $v_n$ of (\ref{vn0}) 
is given by
\begin{eqnarray}
 v_n=  A_{F}'Q^{n-1} {\bf 1},
\mbox{ where }{\bf 1}' = (1,\cdots,1). 
\label{vn1} 
\end{eqnarray}
In terms of  the {\it backward operator} $B$ defined by
$$By_n=y_{n-1},$$
the recurrence relation (\ref{rec}) can be written for $n\geq 0$,
\begin{eqnarray}
u_{n+1}=v_{n+1}+{\cal C}_nu_n\mbox{ where }
{\cal C}_n=I_P+A_F'{\cal B}_n{\bf 1},\
{\cal B}_n=(I_P-Q^nB^n)/(I_P-QB),
\label{hohum} 
\end{eqnarray}
where $I_P$ is the $P\times P$ identity matrix. So ${\cal B}_0=0.$

Alternatively, from (\ref{4}) we have
\begin{eqnarray*}
u_1 &=& 1+A_F'{\bf 1} ,\\
u_2 &=& 1+A_F'(2I+Q){\bf 1} ,\\
u_3 &=& 1+(A_F'{\bf 1} )^2 +A_F'(3I+2Q+Q^2){\bf 1} ,\\
u_4 &=& 1+3(A_F'{\bf 1} )^2 +2(A_F'{\bf 1} )(A_F'Q {\bf 1} ) +A_F'(4I+3Q+2Q^2+Q^3){\bf 1} ,
\end{eqnarray*}
and so on.
Neither solution is satisfactory for large $n$.

{\bf The idempotent case.}
We shall see that
 $Q$ frequently has the form
\begin{eqnarray}
Q=\theta J \mbox{ where }\theta \mbox{ is scalar and }J^2=J.\label{J} 
\end{eqnarray}
That is, $J$ is idempotent with eigenvalues $0,1$. By (\ref{vn1}) 
\begin{eqnarray}
v_1=A_F'{\bf 1},\
 v_{n}= \theta^{n-1}d \mbox{ for }n\geq 2 \mbox{ where } d=A_F'J{\bf 1}. 
\label{vn'}
\end{eqnarray}
This is just (\ref{vn})
with $I=1,\ n_0=2,\ \nu_1=\theta,\ \beta_1=d/\theta$.
By (\ref{un2}) a solution is
$$u_n = \sum_{j=1}^3  \gamma_j \delta_j^n,\ n\geq 3$$
where $\{\delta_j,\ j=1,2,3\}$ are the roots of
$$\delta^3-(\theta+1)\delta^2+(\theta-v_1)\delta+(v_1-d)\theta=0$$
and for $A_2, Q_2$ of (\ref{Q0}),
$$(\gamma_1,\gamma_2,\gamma_3)'=A_2^{-1}Q_2.$$
An explicit solution to a cubic is given in Section 3.8.2 p17 of
 Abramowitz and Stegun (1964).
\begin{example} 
  Suppose that $e_1$ takes only two values, say 0 and 1. 
Then the observations take the
values $0, 1, \rho, 1+\rho$ 
and
\begin{eqnarray*}
Q =\mbox{sign}(\rho)\
 \begin{pmatrix} I(0\leq x/\rho )p_1, & I(0\leq (x-1)/\rho )p_2\\
 I(1\leq x/\rho )p_1, & I(1\leq (x-1)/\rho )p_2
\end{pmatrix} 
.
\end{eqnarray*}
The possible values of $Q$ are $\pm Q_i,\ 1\leq i\leq 8,$ where
\begin{eqnarray*}
Q_1 &=& {p_1p_2\choose p_1p_2},\
Q_2={p_1p_2\choose 0 p_2},\
Q_3={p_1p_2\choose 00},\
Q_4={0p_2\choose 00},\\
Q_5 &=& {0p_2\choose 0p_2},\
Q_6={p_10\choose 00},\
Q_7={p_1 0\choose p_1 0},\
Q_8={p_1p_2\choose  p_1 0}.
\end{eqnarray*}
 For $i=1,3,5,6,7$ and $Q=Q_i$, 
 (\ref{J}) holds with 
$\theta=\theta_i,\ J_i=Q_i/\theta_i,\
\theta_1=1,\ \theta_3=p_1,\ \theta_5=p_2,\
\theta_6=\theta_7=p_1.$ Also (\ref{J}) holds with 
$\theta=0$ for $Q_4$.

There are four cases of $\rho$ to consider.

{\rm The case} ${\rho \leq -1}:$ The observations take the values 
 $ \rho< 1+\rho<0<1.$ 
$Q$ changes at these values of $x$.\\
As $x$ increases through $x\leq \rho , \rho <x\leq 1+\rho ,1+\rho <x\leq 0, 0<x\leq 1$ and $1<x$,
$Q$ changes from $-Q_1$ to $-Q_2$ to $-Q_3$ to $-Q_4$ to 0.

{\rm The case} ${-1<\rho <0}:$ The the observations take the values 
$ \rho<0< 1+\rho<1.$ 
As $x$ increases through $x\leq \rho , \rho <x\leq 0,\ 0<x\leq 1+\rho ,
1+\rho <x\leq 1$ and $1<x$,
$Q$ changes from $-Q_1$ to $-Q_2$ to $-Q_5$ to $-Q_4$ to 0.

{\rm The case} ${0<\rho <1}:$ the observations take the values 
$0< \rho<1< 1+\rho.$ 
As $x$ increases through $x< 0 ,0\leq x<\rho,\ \rho \leq x< 1, 
1\leq x<1+\rho ,
1+\rho \leq x,$ 
$Q$ changes from 0 to $Q_6$ to $Q_7$ to $Q_8$ to $Q_1$.

{\rm The case} ${1\leq\rho}:$ The 
 observations take the values
 $0<1< \rho< 1+\rho.$ 
As $x$ increases through $x\leq 0 ,0\leq x<1,\ 1\leq x<\rho,\ \rho \leq x< 1+\rho, 
1+\rho \leq x$,
$Q$ changes from 0 to $Q_6$ to $Q_3$ to $Q_8$ to $Q_1.$

 Consider {\rm the case} ${0<\rho <1}.$ So $A_{iF}/p_i$ jumps from 0 to $p_1$
to 1 at $x_i+\rho x_1=x_i$ and  $x_i+\rho x_2=x_i+\rho$.
Then there are 5 ranges of $x$ to consider.

$x<0\Rightarrow Q=0$, (\ref{J}) holds with $J=0, \theta=0, 
A_F=0, d=0, u_n=
v_n=0$ for $n\geq 1.$

$0\leq x<\rho\Rightarrow$ (\ref{J}) holds with $Q=Q_6,\ \theta=p_1,
J={10\choose 00}, 
A_F=p_1p, d=p_1^2,
 u_n=v_n=p_1^{n+1}$ for $n\geq 1.$ 

$\rho\leq x<1\Rightarrow$ (\ref{J}) holds with $Q=Q_7,\ \theta=p_1,
J={10\choose 10},
A_F=p_1{1\choose 0}, d=p_1^2,
  u_n=v_n=p_1^{n}$ for $n\geq 1.$ 

$1\leq x<1+\rho\Rightarrow Q=Q_8,\ A_F=p_1{1\choose p_2},$
 but (\ref{J}) does not hold.

$1+\rho\leq x\Rightarrow $ (\ref{J}) holds with $\theta=1,
Q=J=Q_1, 
A_F=p, d=1,\  u_n=v_n=1$ for $n\geq 0.$ 

 Now suppose that ${\rho <0}.$
Then for $i=1,3,5,6,7,\ Q=-Q_i=
\theta_iJ_i$ where $\theta_1=-1,\ \theta_3=-p_1,\ \theta_5=-p_2,\ \theta_6=\theta_7=-p_1$.
Again, this deals with all cases except for $Q_2, Q_4, Q_8$. 

Also $Q_4^2=0$
so that for $Q=\pm Q_4,\ v_n=0$ for $n\geq 2$. For $1+\rho<x\leq 1, A_F=(0,-p_1p_2)', v_1=-p_1p_2, V(t)=1-p_1p_2t.$ So by (\ref{*}), $tU(t)=D^{-1}-1$
where $D=1-tV(t)=(1-p_1t)(1-p_2t)$, giving
$$u_n=(p_1^{n+2}-p_2^{n+2})/(p_1-p_2)$$ for $p_1\neq 1/2$. So for $p_1= 1/2,\
u_n=(n/2+1)2^{-n}.$
This illustrates our second and most general method of solution, the use of
 (\ref{*}).

Finally, the cases $Q_2$ and $Q_8$ can be dealt with by the following method.
\end{example} 
Our third solution is in terms of
 the eigenvalues and left and right eigenvectors of 
$Q$, say $\{\nu_i,\ l_i,\ r_i:\ 1\leq i\leq  P\}$. 

 {\bf The case of diagonal Jordan form (for example distinct eigenvalues).}\\
In this case  the $P\times P$ matrix $Q$ has Jordan canonical form
$$Q=R\Lambda R^{-1}=R\Lambda L'=\sum_{i=1}^P\nu_i r_il_i',\ RL'=I_P, 
\Lambda=diag(\nu_1,\cdots, \nu_P),\ L=(l_1,\cdots,l_P),\ R= (r_1,\cdots,r_P).$$
Then by (\ref{vn1})
\begin{eqnarray}
Q^n &=& L'\Lambda^n R=\sum_{i=1}^P\nu_i^n r_il_i'\mbox{ for }n\geq 0,\nonumber\\
{\cal K}^{n}H(y) &=&   \sum_{i=1}^P b_{iH}(y) \nu_i^{n-1}\mbox{ for }n\geq 1
\mbox{ where } b_{iH}(y)= (A_{H}'r_i) \ (l_i'q(y)),
\label{any}\\
v_n &=& \sum_{i=1}^P \beta_i \nu_i^{n-1} \mbox{ for }n\geq 1
\mbox{ where } 
\beta_i= b_{iF}(\infty)=(A_{F}'r_i)\ (l_i'{\bf 1}). \label{vn3}
\end{eqnarray}
So (\ref{vn}) holds with $I=P,\ n_0=1$.
So for $n\geq 1$, if $\rho>0$, then $u_n=v_n$ of (\ref{vn3}),
and by (\ref{un2}), if $\rho<0$, then
$$u_n = \sum_{j=1}^{P+1} \gamma_j \delta_j^n$$
where
 $\{\delta_j\}$ are 
the roots of
$$
 \sum_{k=1}^P \beta_k\nu_k/(\delta-\nu_k)=\delta-1
$$
and  $\gamma $ is given by (\ref{gam}).
So this method requires computing the left and right eigenvectors of Q for
its non-zero eigenvalues. In rare cases  $Q$ is symmetric so that $L=R.$

 One can show that this method agrees with the idempotent method
 when are both applicable.
\begin{example} 
Let us reconsider the previous example.

Firstly, {\rm suppose that} $0<\rho<1 $.
We consider 3 cases.

{\rm The case}
 $0\leq x<\rho$:  Then $Q=Q_6 $  has eigenvalues $p_1,0$.
For $\nu=p_1, \ l=r={1\choose 0}$. So for $n\geq 1,\ Q^n=p_1^n ll'$
in agreement with the idempotent method.

{\rm The case} $\rho\leq x<1$:  Then $Q=Q_7$ has eigenvalues $0,p_1$.
For $\nu=p_1$, we can take $l={1\choose 0},\ r={1\choose 1}$. So for $n\geq 1,\ 
Q^n= lr' p_1^n$
in agreement with the idempotent method.

{\rm The case} $1\leq x<1+\rho$:  Then $Q=Q_8$  has eigenvalues 
satisfying $\nu^2-p_1\nu-p_1p_2=0$ so that $2\nu_i=p_1\pm(p_1^2+4p_1p_2)^{1/2}$.
 Take $r_i={\nu_i\choose p_1},\
l_i={\nu_i\choose p_2}/c_i$ where $c_i=p_1(\nu_i+2p_2)$. Using 
$\nu_1+\nu_2=p_1,\ \nu_1\nu_2=-p_1p_2$, we obtain
$$Q^n=\sum_{i=1}^2 \nu_i^n 
\begin{pmatrix}
p_1(\nu_i+p_2) & p_2\nu_i\\
 p_1\nu_i &p_1p_2
\end{pmatrix} .$$
Also $A_F'=p_1(1, p_2)$. So one obtains
$$v_n=\sum_{i=1}^2 \nu_i^{n-1}\beta_i\mbox{ for }n\geq 1\mbox{ where }
\beta_i=a_i/(\nu_i+2p_2),\  a_i=(1+p_1p_2)\nu_i+p_1p_2(1+p_2).$$
Secondly, {\rm suppose that} $-1<\rho<x\leq 0 $.
Then $A_F'=-(p_1^2,p_2)$.
Suppose that $p_1\neq 1/2.$
Then $Q=-Q_2,\ Q_2=R\Lambda L'$ where
$\Lambda=diag(p_1,p_2),$ 
\begin{eqnarray*}
R = \begin{pmatrix}
1 &p_2\\ \ 0 & p_2-p_1
\end{pmatrix},\
 L'=R^{-1}=\begin{pmatrix}
 1 & -p_2(p_2-p_1))^{-1} \\ 0 & (p_2-p_1)^{-1}
\end{pmatrix},\
(-Q)^n = R\Lambda^n L'=\begin{pmatrix}
p_1^n & p_2a_n\\ 0 & p_2^n
\end{pmatrix}
\end{eqnarray*}
where  $a_n=(p_1^n-p_2^n)/(p_1-p_2).$
So by (\ref{vn1}), $v_n=(-1)^{n}a_{n+2}.$
So $1+tU(t)=(1-tV(t))^{-1}=(1-\nu_1 t)(1-\nu_2 t)=1+t+p_1p_2t^2$ giving
$u_1=p_1p_2,\ u_n=0$ for $n\geq 2$.
\\ If $p_1=1/2$, then by a limiting argument,  $v_n=(-2)^{-n}(n+2)/2$
for $n\geq 0$, $u_1=1/4,\ u_n=0$ for $n\geq 2$. 
\end{example} 
\begin{example} 
  Suppose that $e_1$ takes the 3 values 0, 1, 2. So the observations take the
values $0, 1,2, \rho, 1+\rho, 2+\rho, 2\rho, 1+2\rho, 2+2\rho$ 
and
\begin{eqnarray*}
Q =\mbox{sign}(\rho)\
 \begin{pmatrix} I(0\leq x/\rho )p_1, & I(0\leq (x-1)/\rho )p_2
 & I(0\leq (x-2)/\rho )p_3
\\
 I(1\leq x/\rho )p_1, & I(1\leq (x-1)/\rho )p_2
 & I(1\leq (x-2)/\rho )p_3
\\
 I(2\leq x/\rho )p_1, & I(2\leq (x-1)/\rho )p_2
 & I(2\leq (x-2)/\rho )p_3
\end{pmatrix} 
.
\end{eqnarray*}
Suppose that
 $0<\rho<1/2.$ Then
$Q$ changes each time $x$ crosses one of the nine values
$0<\rho <2\rho <1<1+\rho <1+2\rho <2<2+\rho <2+2\rho. $
So we need to consider ten cases, six of them idempotent.
The possible values of $Q$ are $\pm Q_i, 0\leq i\leq 9$ 
where
$Q_0=0,$
\begin{eqnarray*}
Q_1 &=&p_1
 \begin{pmatrix}1 & 0 & 0\\
 0 & 0 & 0\\
 0 & 0 & 0
\end{pmatrix}
,\
Q_2=2p_1
 \begin{pmatrix}1 & 0 & 0\\
 1 & 0 & 0\\
 0 & 0 & 0
\end{pmatrix}/2
,\
Q_3 =p_3
 \begin{pmatrix}1 & 0 & 0\\
 1 & 0 & 0\\
 1 & 0 & 0
\end{pmatrix}
,\\
Q_4 &=&
 \begin{pmatrix}p_1 & p_2 & 0\\
 p_1 & 0 & 0\\
 p_1 & 0 & 0
\end{pmatrix}
,\
Q_5 =
 \begin{pmatrix}p_1 & p_2 & 0\\
 p_1 & p_2 & 0\\
 p_1 & 0 & 0
\end{pmatrix} 
,\
Q_6 =
 \begin{pmatrix}p_1 & p_2 & 0\\p_1 & p_2 & 0\\p_1 & p_2 & 0
\end{pmatrix} 
,\\
Q_7 &=&
 \begin{pmatrix}p_1 & p_2 & p_3\\p_1 & p_2 & 0\\p_1 & p_2 & 0
\end{pmatrix} 
,\
Q_8 =
 \begin{pmatrix}p_1 & p_2 & p_3\\p_1 & p_2 & p_3\\p_1 & p_2 & 0
\end{pmatrix} 
,\
Q_9 =
 \begin{pmatrix}p_1 & p_2 & p_3\\p_1 & p_2 & p_3\\p_1 & p_2 & p_3
\end{pmatrix} 
.
\end{eqnarray*}
Also $Q_0$ and $J_i=Q_i/\theta_i$ are idempotent for $i=1,2,3,6,9$ where 
$$\theta_1= \theta_3=p_1,\ \theta_2=2p_1,\ \theta_6=p_1+p_2,\ \theta_9=1.$$

 {\rm Case 1:} $x<0\Rightarrow Q=0,\ A_F=0,\ v_n=0$ for $n\geq 1.$

 {\rm Case 2:} $0\leq x<\rho \Rightarrow Q=Q_1,\ A_F'=(p_1^2,0,0),\ d=p_1^,\ 
v_n=p_1^{n+1}\mbox{ for }n\geq 1.$

 {\rm Case 3:} $\rho \leq x<2\rho \Rightarrow Q=Q_2,\ A_F'=(p_1(p_1+p_2),0,0),\ 
d=p_1(p_1+p_2)/2,$
$$v_1=p_1(p_1+p_2),\ v_n=(2p_1)^{n-1}d\mbox{ for }n\geq 2.$$

 {\rm Case 4:} $2\rho \leq x<1\Rightarrow Q=Q_3,\ A_F'=(p_1,0,0),\ 
v_n=p_1^{n}\mbox{ for }n\geq 1.$ 

 {\rm Case 5:} $1\leq x<1+\rho \Rightarrow Q=Q_4$. $Q_4$ has distinct eigenvalues 
$$0,\nu_2=(p_1+\varepsilon)/2,\nu_3=(p_1-\varepsilon)/2\mbox{ where }
\varepsilon=(p_1^2+4p_1p_2)^{1/2}.$$
Also 
\begin{eqnarray*}
A_F' &=& (p_1,p_1p_2,0),\\
 2\varepsilon \ r_2' &=& (p_1+\varepsilon, 2p_1, 2p_1)=B_\varepsilon\mbox{ say}, 
\ 2p_1 \ l_2'=(2p_1, -p_1+\varepsilon,0)=C_\varepsilon\mbox{ say},\\
 2\varepsilon \ r_3' &=& -B_{-\varepsilon}, 
\ 2p_1 \ l_3'=C_{-\varepsilon}.
\end{eqnarray*}
So by (\ref {vn3}), for $n\geq 1,\ v_n=\sum_{i=2}^3 \beta_i \nu_i^{n-1}$
where 
$$ \beta_2=p_1b_\varepsilon c_\varepsilon/(4\varepsilon),\
\beta_3=-p_1b_{-\varepsilon}c_{-\varepsilon}/(4\varepsilon)
\mbox{ where }b_\varepsilon=p_1+2p_1p_2+\varepsilon,\
c_\varepsilon=2p_1-p_1p_2+p_2\varepsilon
.$$
 {\rm Case 6:} $1+\rho \leq x<1+2\rho \Rightarrow Q=Q_5$. 
This is the only example here where the general Jordan form is needed.

 {\rm Case 7:} $1+2\rho \leq x<2\Rightarrow Q=Q_6={\bf 1} l'=\theta_6 J_6$ say,
where $l'=(p_1,p_2,0), \theta_6=p_1+p_2,\ J_6^2=J_6$.\\
 Also $A_F'=(p_1,p_2,0)$. 
So by (\ref{vn'}),
$v_n= (p_1+p_2)^n$ for $n\geq 1.$ 

 {\rm Case 8:} $2 \leq x<2+\rho \Rightarrow Q=Q_7.\ Q_7$
 has distinct eigenvalues 
$$0,\nu_2=(p_1+p_2+\varepsilon)/2,\nu_3=(p_1+p_2-\varepsilon)/2\mbox{ where }
\varepsilon=((p_1+p_2)^2+4p_1p_3)^{1/2}.$$
 Also 
\begin{eqnarray*}
A_F' &=& (p_1,p_2,p_1p_3), \\ 
 2p_1\varepsilon \ r_2' &=& (a_\varepsilon,b_\varepsilon,b_\varepsilon)
\mbox{ where }a_\varepsilon=p_1p_2+2p_1p_3+p_2^2-p_2\varepsilon, \
b_\varepsilon=p_1(p_1+p_2-\varepsilon),
\\ 2p_1p_3 \ l_2' &=&(p_1c_\varepsilon,p_2c_\varepsilon, 2p_1p_3)\mbox{ where }
c_\varepsilon=p_1+p_2+\varepsilon,\\
2p_1\varepsilon \ r_3' &=& (-a_{-\varepsilon},b_{-\varepsilon},b_{-\varepsilon}),
\ 2p_1p_3 \ l_3' =(p_1c_{-\varepsilon},\ p_2c_{-\varepsilon},\ 2p_1p_3),
\end{eqnarray*}
so that (\ref {vn3}) for $n\geq 1,\ v_n=\sum_{i=2}^3 \beta_i \nu_i^{n-1}$
where
\begin{eqnarray*}
 \beta_2 &=& B_{2\varepsilon}C_{\varepsilon}/(4p_1^2p_3 \varepsilon),\
 \beta_3=B_{3 -\varepsilon}C_{-\varepsilon}/(4p_1^2p_3 \varepsilon),\\
\mbox{where }  B_{2\varepsilon} &=& p_1a_{\varepsilon}+(p_2+p_1p_3)b_{\varepsilon},\
 B_{3 -\varepsilon} = -p_1a_{-\varepsilon}+(p_2+p_1p_3)b_{-\varepsilon},\\
C_{\varepsilon} &=&(p_1+p_2)c_{\varepsilon}+2p_1p_3.
\end{eqnarray*} 
 {\rm Case 9:} $2+\rho \leq x<2+2\rho \Rightarrow Q=Q_8.\ Q_8$
  has distinct eigenvalues 
$$0,\nu_2=(p_1+p_2+\varepsilon)/2,\ \nu_3=(p_1+p_2-\varepsilon)/2 \mbox{ where }
\varepsilon=\delta^{1/2},\
\delta=(p_1+p_2)(p_1+p_2+4p_3=1+3p_3).$$
By (\ref {vn3}) for $n\geq 1,\ v_n=\sum_{i=2}^3 \beta_i \nu_i^{n-1}$
where 
\begin{eqnarray*}
A_F' &=& (p_1,p_2,p_3(p_1+p_2)),\\ 
 2(p_1+p_2)\ \varepsilon r_2' &=& (p_1(p_1+p_2+\varepsilon),\ p_1(p_1+p_2+\varepsilon),\
p_1)=B_\varepsilon\mbox{ say} ,\\
 2p_1\  l_2' &=& ( 2p_1,\ 2p_2,\ -p_1-p_2+\varepsilon)=C_\varepsilon\mbox{ say}
,\\
 2(p_1+p_2)\varepsilon \ r_3' &=& -B_{-\varepsilon},\
 2p_1\  l_3' = C_{-\varepsilon}.
\end{eqnarray*}
 {\rm Case 10:} $2+2\rho \leq x\Rightarrow Q=Q_9$. As noted, (\ref{J} holds
with $\theta_9=1.$ Also $Q_9{\bf 1}={\bf 1},\ A_F'=(0,0,p_3), d=p_3$. So
by (\ref{vn'}), $v_n=p_3$ for $n\geq 1.$

This leaves only $Q_5$ to deal. It will be dealt with by the following method.
\end{example} 
Our third and general solution is in terms of
 the eigenvalues and left and right {\it generalized} eigenvectors of 
$Q$, say $\{\nu_i,\ l_i,\ r_i:\ 1\leq i\leq  P\}$. 

 {\bf The general Jordan form.}\\

The general  Jordan canonical form for a $q\times q$ matrix $Q$ is
\begin{eqnarray}
Q &=& R\Lambda R^{-1}=R\Lambda L'=\sum_{i=1}^r R_iJ_{m_i}(\nu_i)L_i',
\label{gen}
\\
\mbox{where } RL' &=& I, \ L=(L_1,\cdots,L_r),\ R= (R_1,\cdots,R_r),\
\Lambda=\mbox{diag}(J_{m_1}(\nu_1),\cdots, J_{m_r}(\nu_r)), \nonumber
\end{eqnarray}
and $L_i, R_i$ are $q\times m_i$, 
$$J_m(\nu)=\nu I_m+U_m,$$ 
and $U_m$ is the
$m\times m$ matrix with zeros everywhere except for ones on the diagonal above
the leading diagonal: $(U_m)_{ij}=\delta_{i,j-1}$:
$$J_1(\nu)=\nu,\ J_2(\nu)={\nu 1\choose 0 \nu},\
 J_3(\nu)= 
\begin{pmatrix}\nu & 1 & 0\\0 & \nu & 1\\0 & 1 & \nu
\end{pmatrix},\cdots 
$$
$J_m(\nu)$ has only one right eigenvector. 
The $i$th block in $QR=R\Lambda$ is $QR_i=R_iJ_{m_i}(\nu_i)$. Taking its $j$th
column gives
$$Qr_{ij}=\nu_i r_{ij}+ r_{i,j+1}, j=1,\cdots,m_i,\mbox{ where }r_{i,m_i+1}={\bf 0} $$
and $r_{ij}$ is a $q$-vector. So one first computes the right eigenvector
$r_{i,m_i}$ and then the {\it generalized} eigenvectors
 $r_{i,m_i-1},\cdots, r_{i1}$ recursively, the Jordan chain.
For $n\geq 0$, the $n$th power of $Q$ is
\begin{eqnarray}
J_{m}(\nu)^n  &=&
 \sum_{k=0}^{\min(n,m-1)} {n\choose k} \nu^{n-k} U_m^k
 \label{jdn},\\
Q^n  &=&
 R\Lambda^n R^{-1}
=\sum_{i=1}^r R_i \ J_{m_i}(\nu_i)^n \ L_i'
=\sum_{i=1}^r \sum_{k=0}^{\min(n,m_i-1)} {n\choose k} \nu_i^{n-k} W_{ik}
 \label{qn}\\
\mbox{where } W_{ik} &=& R_iU_{m_i}^kL_i'.\nonumber
\end{eqnarray}
where $ U_m^k$ is  the
$m\times m$ matrix with zeros everywhere except for ones on the $k$th
super-diagonal: $(U_m^k)_{ij}=\delta_{i,j-k}$.
So $U_m^m=0$. 
So $Q^n$ is a matrix polynomial in $n$ of degree $m-1$ where $m=\max_{i=1}^r m_i$, and by (\ref{vn1}),
\begin{eqnarray}
v_n=\sum_{i=1}^r \sum_{k=0}^{\min(n,m_i)-1} {n-1\choose k} \nu_i^{n-1-k} w_{ik}
\mbox{ for } n\geq 1 \mbox{ where }  w_{ik}= A_F'W_{ik}{\bf 1}.\label{vnm}
\end{eqnarray}
For diagonal Jordan form this reduces to (\ref{vn}).
This level of generality is not needed if the eigenvalues of the non-diagonal
Jordan blocks are zero, since we can rewrite (\ref{qn}) and (\ref{vnm}) as
\begin{eqnarray}
Q^n  &=&
\sum_{1\leq i\leq r,\ \nu_i\neq 0}
\sum_{k=0}^{\min(n,m_i-1)} {n\choose k} \nu_i^{n-k} W_{ik}
+\sum_{1\leq i\leq r,\ \nu_i= 0}I(n<m_i) W_{in},\  n\geq 0,
 \label{qn0}\\
v_n &=&
\sum_{1\leq i\leq r,\ \nu_i\neq 0} \sum_{k=0}^{\min(n,m_i)-1}
 {n-1\choose k} \nu_i^{n-1-k} w_{ik} 
+\sum_{1\leq i\leq r,\ \nu_i= 0}I(n\leq m_i) w_{i,n-1}, \ n\geq 1
.\label{vnm0}
\end{eqnarray}
The solution for $u_n$ is a weighted sum of $n$th powers where the weights
are constants or polynomials in $n$: see   Withers and Nadarajah (2009b)
for details.
\begin{example} 
This continues Case 6 of Example 3.3.
Set $q_i=p_i/(p_1+p_2).$ By Appendix A, (\ref{gen}) holds for $Q=Q_5$
with $r=2, m_1=2,\ m_2=1, \nu_1=0,\ \nu_2=p_1+p_2,$
\begin{eqnarray*}
R_1 =
 \begin{pmatrix}0 & q_2 \\
 0 & -q_1 \\
 q_2 & -q_1^2 
\end{pmatrix} ,
\
R_2 = q_1
 \begin{pmatrix} 1\\
 1\\
  q_1
\end{pmatrix} ,
\
L_1'= \begin{pmatrix}0 & -p_1/p_2 & 1+p_1/p_2 \\
 1 & -1 & 0
\end{pmatrix},
\
L_2'= \begin{pmatrix} 1 & p_2/p_1 & 0
\end{pmatrix}.
\end{eqnarray*}
Also $J_2(0)^n=0$ for $n\geq 2.$ So 
$$Q^n=(p_1+p_2)^nR_2L_2'=(p_1+p_2)^{n-1}
\begin{pmatrix}p_1 & p_2 & 0\\ p_1 & p_2 & 0\\
 p_1q_1 & p_2q_1 & 0
\end{pmatrix}
\mbox{ for } n\geq 2.$$
Also $A_F'=(p_1,p_2(p_1+p_2),0).$ So by (\ref{vn1}),
$$v_n=
\nu^{n-1}v_1\mbox{ for }n\geq 1
\mbox{ where }\nu=p_1+p_2,\ v_1=p_1+p_2(p_1+p_2).$$
As noted, 
this is the only example here where the general Jordan form is needed.
\\(It was implicit in  the 2nd part of Example 3.2 for the case $p_1=1/2$,
but was bypassed by the limiting argument.) So $V(t)=1+v_1t/(1-\nu t),\
1-tV(t)=L/(1-\nu t),\ L=(1-t)(1-\nu t)-v_1t^2=(1-tt_1)(1-tt_2),\
t_k=(\nu+1\pm \delta^{1/2})/2,\
\delta=(\nu+1)^2-4(\nu-v_1)=p_3^3+4v_1,\
1+tU(t)=(1-tV(t))^{-1}=(1-\nu t)/L=\sum_{k=1}^2c_k/(1-tt_k),\
c_k=(1-\nu/t_k)(1-t_{3-k}/t_k)$ giving for $\rho<0$,
$$u_{n-1}=\sum_{k=1}^2c_k t_k^n,\ n\geq 1.$$
So although $m_1=2$, a sum of powers solution still holds.
\end{example}
As noted for the diagonal form,
$Q$ and its
eigenvectors only change
value when $x$ crosses one of the possible observation values, or equivalently when
$\rho$ crosses one of the set  $\{(x-x_j)/ x_i,\ 1\leq i,j\leq P\}$.
If $x$ is replaced by $x_n$, then again $Q$ and its eigenvalues do not depend on
$n$ or $x_n$ except through its range. 

For more on Jordan forms, see for example 
{\tt http://en.wikipedia.org/wiki/Jordan\_normal\_form}
\begin{note} 
The singular values of $J_m(\nu)$ needed for its singular value
decomposition, SVD,
 are quite different from its eigenvalues, which are all $\nu$.
The SVD of $Q$ gives an alternative form for its inverse, but is of no use
in computing $Q^n$.
\end{note}

\section{A solution for $G_n(y)$.}
\setcounter{equation}{0}
A solution for $G_n(y)= Prob(M_n\leq x,\ e_n\leq y)$
may be of interest.

If $\rho>0$ then a solution is given by (\ref{Gpos}):
$$G_n(y)=a_n(y)\mbox{ where }a_n(y)={\cal K}^n  F(y),\ n\geq 1,
$$
Suppose that $Q$ of (\ref{Q})
has diagonal Jordan form. Then by (\ref{any}),
$$a_n(y)= A_{F}'Q^{n-1} q(y) =\sum_{i=1}^P b_{iF}(y) \nu_i^{n-1}
\mbox{ where } b_{iF}(y)= (A_{F}'r_i) \ (l_i'q(y)) \mbox{ for } n\geq 1.$$
Now suppose that $\rho<0$. Then
 a solution  is given by (\ref{2.4a}):
$$G_{n+1}(y)
 = a_n(y) \otimes u_n +a_{n+1}(y),\ n\geq 0.$$
Also by Section 2, (\ref{vn}) holds with $n_0=1$:
$$v_{n}= \sum_{j=1}^I \beta_j \nu_j^{n-1} 
\mbox{ for } n\geq 1
\mbox{ where }\beta_j= b_{jF}(\infty)=(A_{F}'r_i) \ (l_i{\bf 1}),$$
so that by (\ref{un2})
$$u_n = \sum_{j=1}^{I'} \gamma_j \delta_j^n \mbox{ for } n\geq 1.$$
Substituting we obtain using (\ref{un2}),
\begin{eqnarray}
G_{n+1}(y) &= & 2\sum_{i=1}^P b_{iF}(y) \nu_i^{n-1}
-\sum_{j=1}^{I'} \gamma_j c_j(y) \delta_j^{n+1} \mbox{ for } n\geq 1
\\
\mbox{where }c_j(y) &=& \sum_{i=1}^P b_{iF}(y)
\nu_i^{-1}(\nu_i-\delta_j)^{-1}.
\label{gnyfinal}
\end{eqnarray}
So its behaviour for large $n$ is determined by the $\nu_i$ or $\delta_j$ of
largest modulus.
\appendix
\section{Appendix: a MAPLE program to find the Jordan form.}%
\setcounter{equation}{0}

Here is the MAPLE program used to work out the Jordan form for $Q=Q_5/p_1$ in 
Example 3.4. It can easily be adapted to other examples.
We set $c=p_2/p_1,\ M=L'$.
\begin{verbatim}
with(LinearAlgebra);
 Q:=Matrix(3,3,[[1,c,0],[1,c,0],[1,0,0]]);
JordanForm(Q);
R:= JordanForm(Q, output='Q') ; 
M:=R^(-1); 
J:=simplify(M.Q.R);
quit;
\end{verbatim}
The last line is just to confirm that  $Q=RJM$.

\end{document}